\documentclass[fleqn,usenatbib]{mnras}

\usepackage{newtxtext,newtxmath}

\usepackage[T1]{fontenc}
\defcitealias{Kumar2022}{KMR22}

\DeclareRobustCommand{\VAN}[3]{#2}
\let\VANthebibliography\thebibliography
\def\thebibliography{\DeclareRobustCommand{\VAN}[3]{##3}\VANthebibliography}

\usepackage{graphicx}	
\usepackage{amsmath}	

\usepackage{enumerate}

\newcommand{\bc}{}

\newcommand{\msunh}{${h^{-1}\rm M_\odot}~$}
\newcommand{\msunhh}{{h^{-2}\rm M_\odot}}

\newcommand{\mpch}{{h^{-1}\rm Mpc}}
\newcommand{\mpcH}{${h^{-1}\rm Mpc}~$}

\newcommand{\vcutsymbol}{{\ooalign{\hfil$\vee$\hfil\cr\kern0.08em--\hfil\cr}}}

\newcommand{\laM}{$\lambda ~$}

\title[Over-abundance of orphans in {\sc  UniverseMachine}]{Over-abundance of orphan galaxies in the {\sc  UniverseMachine}}

\author[Kumar et al.]{
Amit Kumar,$^{1}$\thanks{E-mail: amitk@iucaa.in}
Surhud More,$^{1,2}$\thanks{E-mail: surhud@iucaa.in}
Tomomi Sunayama$^{3,4}$
\\
$^{1}$ Inter-University Centre for Astronomy and Astrophysics, Post bag 4, Ganeshkhind, Pune 411007, India\\
$^{2}$ Kavli Institute for the Physics and Mathematics of the Universe (WPI), 5-1-5 Kashiwanoha, 2778583, Japan\\
$^{3}$Steward Observatory, University of Arizona, Tucson, AZ, 85719, U.S.A\\
$^{4}$Kobayashi-Maskawa Institute for the Origin of Particles and the Universe (KMI), Nagoya University, Nagoya, 464-8602, Japan\\
}

\date{Accepted XXX. Received YYY; in original form ZZZ}

\pubyear{2023}

\begin{document}
\label{firstpage}
\pagerange{\pageref{firstpage}--\pageref{lastpage}}
\maketitle

\begin{abstract}
Orphan galaxies that have lost a large fraction of the dark matter subhaloes have often been invoked in semi-analytical as well as empirical models of galaxy formation. We run a mock cluster finder that mimics the optical cluster finding technique of the redMaPPer algorithm on a catalogue of galaxies with quenched star formation from one such empirical model, the {\sc  UniverseMachine}, and obtain the prevalence of orphan galaxies in these clusters as a function of their cluster-centric distance. We compare the fraction of orphan galaxies with the upper limits derived based on our prior observations of the weak lensing signals around satellite galaxies from SDSS redMaPPer clusters. Although the orphan fraction from the {\sc  UniverseMachine} is marginally consistent with the upper limits in the innermost regions of galaxy clusters spanning $[0.1, 0.3] \mpch$, we observe that the orphan fractions substantially violate the upper limits in the outer regions of galaxy clusters beyond $0.3\mpch$. We discuss the reasons, plausible improvements to the model and how observations can be used to constrain such models further.
\end{abstract}


\section{Introduction}
Satellite galaxies in galaxy clusters live in dense environments, and hence are susceptible to various environmental effects such as ram-pressure stripping, galaxy harassment, and tidal disruption of their dark matter subhaloes. These physical processes can alter the characteristics of satellite galaxies, such as removing their hot gas and quenching their star formation \citep{Gunn1972}, disrupting their spiral arms \citep{Moore1996, Moore1998}, and stripping the mass from their dark matter subhaloes mainly from the outskirts \citep{Merritt1983, Giocoli2008}. Hence, these processes can impact the evolution of satellite galaxies in galaxy clusters.

Tidal effects on satellites are stronger near the cluster center (the brightest central galaxy i.e. BCG), and fall off as the separation from the BCG increases \citep{Springel2001, Lucia2004, Gao2004, Zhao2004, Xie2015}. The satellite galaxies also experience dynamical friction, causing their orbits to decay with time and transport them towards the center of the host's potential well \citep{Binney2008}. If dark matter has self-interactions, that will also evaporate some of the dark mass in subhaloes \citep{Spergel2000, Bhattacharyya_2021}. These physical processes affect the dark matter distribution around satellite galaxies. 

One can use cosmological simulations to study the effect of environmental processes on the evolution of satellites in various environments. Collisionless dark matter simulations have shown mass loss due to tidal stripping for subhaloes \citep{Ghinga1998, Gao2004, vandenBosch2005, Nagai2005, Giocoli2008, Xie2015, Rhee2017}. Empirically, one can use weak lensing observations to understand the dark matter distribution around satellites in a dense environment. We can learn about the effects of a dense environment by comparing matter distribution around satellites with the distribution around galaxies having similar properties but residing in the field environment \citep[hereafter KMR22]{Kumar2022}.

In cosmological simulations, the extent to which a satellite survive depends on the mass resolution \citep{Klypin1999}, halo finder \citep{Onions2012}, and the baryonic physics included \citep{Zolotov2012}. There is growing evidence of artificially disrupted subhaloes due to insufficient resolution in numerical simulations \citep[see e.g.,][]{vandenBosch2017, vandenBosch2018, vandenBosch2018b}. The satellites that have lost most of their dark mass, and are no longer resolved by the halo finder are termed as \textit{orphan galaxies}. The primary reason for a satellite to turn into an orphan after infall to the host halo are close peri-centric passages in their orbit around the central galaxy. Such passages can tidally disrupt the mass from the subhalo, and the orphans may even merge with the central galaxy at later stages of their evolution. In simulations, most galaxies that turn into orphan satellites are found within the virialized radius, $R_{\rm vir}$ of the central galaxy \citep{Guo2011}. Orphan galaxies have not been traditionally invoked by most previous empirical models such as abundance matching \citep{Reddick2013} in order to explain galaxy clustering observations. In order to achieve this, subhaloes of a given mass proxy are forced to host satellite galaxies with larger stellar masses than haloes that correspond to the central galaxies. Although some of these conclusions have to be revisited given the pseudo-evolution of halo masses due to changing halo mass definition \citep{DiemerMore2013, More2015}, this is quite contrary to how satellite galaxies are expected to evolve. Satellite galaxies are expected to be more quenched due to environmental effects than field galaxies. 

A plausible solution to such apparent inconsistencies is to invoke the existence of orphan satellite galaxies \citep{Campbell2018, Behroozi2019, DeRose2022}, which do not reside in any identifiable dark matter subhaloes. Apart from numerical issues with simulations, one could expect a fraction of galaxies to outlive their subhaloes. The presence of a galaxy at the centre of dark matter concentrations could make the subhalo more robust to stripping than in the collisionless scenario. Empirical studies that fit the abundance and clustering of galaxies self-consistently and thus often need to invoke substantial fractions of orphan galaxies to match observational data, in particular at the low stellar mass end \citep{Behroozi2019}. 

In our recent analysis \citetalias{Kumar2022}, we looked for the mass distribution around satellite galaxies identified by the redMaPPer \citep{Rykoff2014, Rykoff2016} cluster finding algorithm with the help of weak gravitational lensing from the Year 1 galaxy shape catalogue from the Subaru Hyper Suprime-Cam survey. We compared this mass distribution to that of galaxies that have similar photometric properties but reside in non-cluster environments and used the observed mass difference to put an upper bound on the population of orphan galaxies admissible at different cluster-centric distances. Here in this work, we wish to compare our upper bounds with the prevalence of orphan galaxies in the publicly available galaxy catalogues from the {\sc  UniverseMachine} algorithm. It is well known that the membership of satellite galaxies in optically identified clusters could have a substantial fraction of interloper galaxies, which are not gravitationally bound to the main structure. The presence of such galaxies could result in an incorrect inference of the fraction of orphan galaxies. Therefore, we use a mock optical cluster finding algorithm to mimic projection effects in order to carry out a fair comparison to our results.

\begin{figure}
\includegraphics[width=0.95\columnwidth]{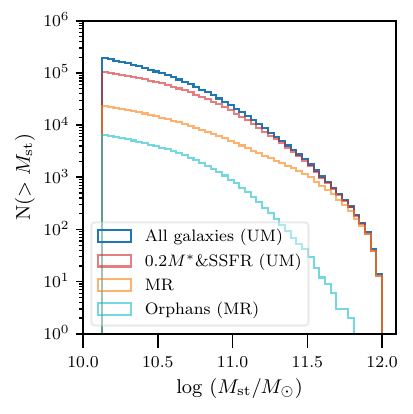}
\caption{The figure depicts the population of galaxies in stellar masses. The blue colour histogram shows the entire galaxy dataset from the {\sc  UniverseMachine} catalogue, and the red colour reflects the same galaxy population subjected to characteristic stellar mass ($M^*$) and specific star formation rate (SSFR) cuts. Similarly, the orange histogram illustrates the implementation of mock-redMapper on the original catalogue, with the orphan population shown with cyan colour. (Refer to the text for definitions).}
\label{fig:galaxy_distribution}
\end{figure}

\section{Data}
\label{sec:data}
In this section, we will briefly discuss the {\sc  UniverseMachine} \citep{Behroozi2019} framework and the galaxy catalogue \footnote{\label{data}\url{http://halos.as.arizona.edu/UniverseMachine/DR1/SFR_ASCII/}, Public DR1} obtained from applying this algorithm to the halo merger trees from the Bolshoi-Planck \citep{Rodriguez2016} simulation.

\subsection{The {\sc  UniverseMachine}} 
The {\sc  UniverseMachine} adopts an empirical approach to model the star formation rates of individual galaxies within their dark matter haloes, given a proxy of the depth of the potential well, its assembly history, and the redshift. The model parameters that describe the dependence of the star formation rate and the fraction of galaxies that are quenched as a function of these variables are constrained using a number of observations that span a wide redshift range up to $z<10$. These observations include the abundances of galaxies, the cosmic star formation rate, the fraction of quenched galaxies as a function of galaxy observables as well as the auto and cross-correlation of star-forming and quenched galaxies. The {\sc  UniverseMachine} forward models these observables by simulating observational errors and biases before comparing the predictions of the model to the observables.

The {\sc  UniverseMachine} uses a MCMC algorithm to sample the posterior distribution for the empirical model, using priors constraints from large observational surveys, finally converging to the best fit position with the help of a gradient descent algorithm. As a by product, the {\sc  UniverseMachine} generates a number of predictions, the stellar-to-halo mass relation as a function of redshift, the star formation histories of galaxies, their correlation functions, as well as infall and quenching statistics for satellite galaxies. The galaxy catalogue from the {\sc  UniverseMachine} also provides various useful representations of the galaxy-dark matter connection for both central and satellite galaxies in addition to the galaxy properties such as their true and observed stellar masses, sizes, and whether these galaxies are orphan galaxies. 

\begin{figure}
    \includegraphics[width=0.9\columnwidth]{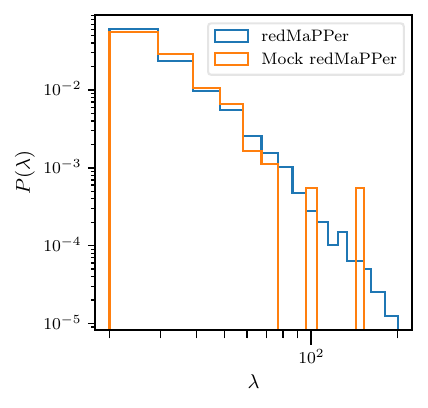}
    \caption{This figure depicts the probability density of clusters plotted against their richness parameter $\lambda$. In this representation, the blue colour corresponds to clusters identified in the redMaPPer catalogue, while the orange colour represents clusters with $\lambda$>20, identified using our mock redMaPPer algorithm applied to the {\sc  UniverseMachine} galaxy catalogue.}
    \label{fig:lambda_distribution}
\end{figure}

We use the publicly available halo and galaxy catalogue at redshift zero obtained from the implementation of the {\sc  UniverseMachine} on halo catalogues from Bolshoi-Plank simulation, which is a collisionless dark matter simulation of a flat-$\Lambda$CDM universe with cosmological parameters $\Omega_{\rm m}$ = 0.307, $\Omega_{\rm b}$ = 0.048, $h$ = 0.678, $n_{\rm s}$ = 0.96 and $\sigma_8$ = 0.823. The simulation evolves $2048^3$ particles each having mass $1.5 \times 10^8$ \msunh in a cubical box of length $250 \mpch$.

\section{Methodology}
\label{sec:methodology}
We apply the cluster finding algorithm of \cite{SunayamaMore2019}, which is designed to approximate the redMaPPer algorithm \citep{Rykoff2014, Rykoff2016}, in order to find clusters in the publicly available {\sc  UniverseMachine} galaxy catalogue. The original redMaPPer algorithm relies on the presence of an overdensity of red galaxies in a region of the sky. The approximately linear relation between the colour of galaxies and the magnitude of quenched galaxies that reside within a galaxy cluster is called the red sequence. The exact colour of galaxies on the red sequence depends on the redshift of galaxies which determines the location of the 4000 \r{A} break in a set of broadband filters. The redMaPPer algorithm fits a quenched galaxy template to identify such red sequence galaxies at various redshifts and in turn identifies overdensities of red galaxies with similar colours that are clustered in the plane of the sky to group them into galaxy clusters. The intrinsic scatter of the red sequence and the photometric uncertainties result in a population of interloper galaxies that masquerade as cluster members.

Instead of mimicking the entire algorithm, we will skip the initial step of identifying galaxies on the red sequence based on their colours, but instead assume that the red sequence technique identifies galaxies with low star formation rates (we assume specific star formation rate $<1.5\times10^{-11} yr^{-1}$). We will assume that the redshifts of such galaxies determined by the red sequence technique will have some inherent uncertainties and can result in grouping of galaxies that are separated by $\pm 50\mpch$. We then follow a procedure similar to the original redMaPPer algorithm to identify clusters. The original algorithm is run on galaxies with luminosities $L \geq 0.2 L_*$, where the luminosity $L_*$ corresponds to the knee of the Schechter function fit to the luminosity function. We use galaxies with stellar masses larger than $0.2M^*$ instead where $M_*=6.7\times10^{10} M_{\odot}$ \citep{Li_white2009}. We have explicitly checked that the knee of the Schechter function in the {\sc  UniverseMachine} catalogues corresponds to this number. In the observational redMaPPer cluster catalogue, such a cut reduces the scatter in the halo mass-richness relation. We construct a filter for galaxies that depends upon the projected distance from any given putative cluster center. Each galaxy obtains a membership probability, $p_{\rm mem}$ to be part of a galaxy cluster. These probabilities are related to the overall richness \laM for the cluster with the following set of implicit equations,
\begin{align}
p_{\rm mem} = \sum{\frac{\lambda~u}{\lambda~u + b}}\,,\,\,\,\,\lambda = \sum{p_{\rm free} ~ p_{\rm mem}} \label{eq:lambda}\,,
\end{align}
where $b$ denotes the expected background galaxy density, and $u$ is a radial spatial filter that corresponds to a projected Navarro, Frenk, and White \citep{NFW1997} profile with a concentration of $5$, truncated smoothly at $R = R_c$, and normalized to unity \citep[see Sec:3 ][]{Rozo2009}. The cut-off radius, $R_c=R_0(\lambda/100)^{\beta}$, where $R_0=1.0$ physical \mpcH and $\beta$ = 0.2, \citep[see Eq.4][]{Rykoff2011}. The value of $p_{\rm free}$ denotes the probability that a galaxy is part of a cluster identified previously.

The cluster finding is performed in an iterative manner. In the beginning of every cluster finding iteration, the parameter $p_{\rm free}$ is initially set to unity for all galaxies. The algorithm starts by assigning richnesses to galaxies rank ordered by their stellar masses, at the beginning of the cluster finding run. The parameter $p_{\rm mem}$ is updated every time a galaxy is assigned to a galaxy cluster, so that progressively it has a lower probability of getting assigned to lower mass galaxy clusters. The only unknown quantity in the above equation is \laM, for which we can solve the equation numerically. The cluster finding is followed by a step which iteratively finds the cluster center based on the luminosity of galaxies, the presence of overdensity of satellite galaxies near it. In successive iterations, the cluster finding is performed by starting from the richest cluster and progressively identifying clusters of lower richnesses. For the first iteration where the richnesses are not available, we carry out an iteration with a fixed cut-off radius of $0.5 \mpch$. At the end of each iteration we assign the central galaxy to be the most massive within the members, and recompute the richness around this central galaxy.

Fig.~\ref{fig:galaxy_distribution} compares the stellar mass distribution of galaxies in the original {\sc  UniverseMachine} catalogue (blue histogram) with the red galaxies on which we run the mock algorithm (shown with the red histogram). The stellar mass distribution of member galaxies is shown with the orange histogram. The fraction of galaxies that are cluster members is larger at the high stellar mass end. In this figure we also show the stellar mass distribution of the orphan galaxies (cyan histogram) which are members of the galaxy clusters identified by the mock algorithm.

\begin{figure}
    \includegraphics[width=0.90\columnwidth]{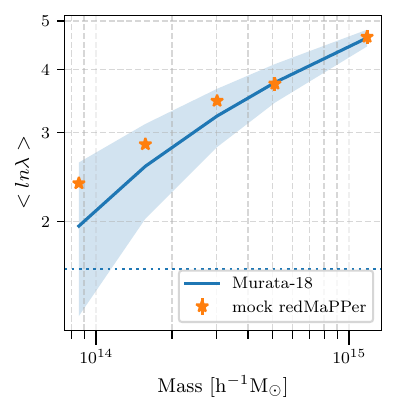}
    \caption{The orange colour represents the mass richness relation for mock redMaPPer clusters classified on the {\sc  UniverseMachine} catalogue, whereas the blue curve corresponds to predictions from \citealt{Murata2018}. Here the mock redMaPPer classifies clumps with $\lambda$>5 as clusters. The dotted blue line in the figure corresponds to this richness threshold. The error bars on {\sc  UniverseMachine} come from  Poisson statistics.}
    \label{fig:mass_richenss}
\end{figure}

In Fig.\ref{fig:lambda_distribution}, we compare the richness distribution of clusters in the original and the mock redMaPPer runs, which are shown with blue and orange colours, respectively. The redMaPPer catalogue is approximately complete for $\lambda>20$ within $0.1< z < 0.33$. Therefore for comparison, we adopt a similar cut in the mock redMaPPer catalogue. We see a broad agreement between the two richness distributions which suggests that our mock algorithm effectively selects clusters in the mock catalogue that are likely consistent with the original redMaPPer catalogue. As a further check, in Fig.\ref{fig:mass_richenss}, we compare the mass-richness relation for all the clusters identified in the mock redMaPPer run with the mass-richness relation derived for redMaPPer clusters in real data.  The orange colour points with errorbars show the dependence of the mean logarithmic richness on the mass of the mock redMaPPer cluster. While computing the mass for any individual mock cluster, we assign the heaviest halo mass corresponding to any cluster member galaxy. The inference of the mass-richness relation based on the weak lensing measurements of clusters in the real redMaPPer catalogue from \cite{Murata2018} are shown as a dotted blue line in the Figure. In this comparison, note that we have included all the clusters with  $\lambda>5$, which is indicated by the dotted blue line in the figure. As \cite{Murata2018}, measure the underlying mass-richness relation we see that the two agree reasonably well at the high mass end. The limiting richness cut of $5$ for mock redMaPPer clusters imply that we start to see deviations at the low richness end. The agreement seen in this figure is non-trivial since the optical selection effects are expected to affect such a mass richness relation, and it goes on to show that the mock algorithm is able to capture such effects to a great degree.

\begin{figure} 
    \includegraphics[width=0.9\columnwidth]{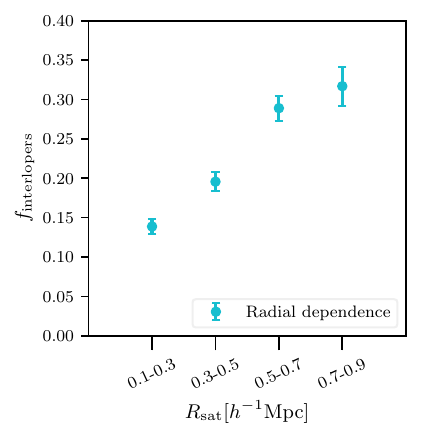}
    \caption{The cyan colour depicts the fraction of interloper satellites at various cluster centric distances as identified in the mock {\sc  UniverseMachine} cluster catalogue.}
    \label{fig:interlopar_fraction}
\end{figure}

\begin{figure}
    \includegraphics[width=0.9\columnwidth]{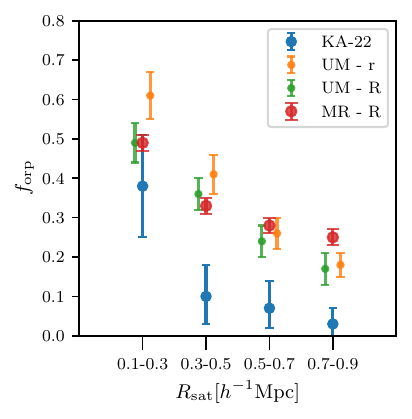}
    \caption{The figure shows the fraction of orphan galaxies that constitute the satellite population inhibiting galaxy clusters at various distances from the host center. The red colour represents the results from satellites defined with Mock-redMaPPer. Orange (green) colour dots represent the orphan prevalence from the original catalogue considering 3-dimensional (projected, i.e. 2-dimensional) separation of the satellites from the host. The blue colour data represents the upper limit admissible on orphan prevalence from observational redMaPPer+HSC analysis by \citetalias{Kumar2022}.}
    \label{fig:orphan_fraction}
\end{figure}

\section{Results and Discussion}

In \citetalias{Kumar2022}, we measured the weak lensing signals around satellite galaxies in real redMaPPer clusters and compared them to the lensing signal around a control sample of photometrically equivalent field galaxies, as a function of the distance of the satellite away from the cluster center. Our results showed a difference between the subhalo masses of the satellite galaxies and the control sample. This difference could be a result of subhalo masses being smaller due to the stripping of material inside the galaxy cluster. The difference allowed us to put an upper limit on the fraction of orphan galaxies in the cluster as a function of the cluster centric radius. These upper limits are shown with blue coloured points with errorbars in Fig.~\ref{fig:orphan_fraction}. The upper limits are less stringent at the innermost radii because the weak lensing signal between the satellite galaxies and the control sample of field galaxies differs by quite a significant amount. These upper limits become more and more stringent at larger distances away from the cluster center and can be at most 10 percent according to our previous observations.

In the {\sc  UniverseMachine} galaxy catalogues, orphan galaxies have been separately tagged and thus identified. First, we compute the fraction of orphan galaxies that we would like to ideally constrain in galaxy clusters, if we were able to measure the 3-dimensional positions of galaxies. For this purpose, for all cluster haloes defined in the {\sc  UniverseMachine} catalogues, we compute the number of member galaxies that satisfy our star formation rate and stellar mass cut and lie within their 3-D halo boundary. We only use clusters where the number of such members is larger than $20$. The orphan fraction as a function of the 3-D cluster-centric distance is shown as orange colour points with errors. The orphan fractions in the {\sc  UniverseMachine} galaxy clusters are as high as $60\%$ in the innermost $[0.1, 0.3] \mpch$. Although the values of these orphan fractions decline with distance away from the cluster center, they remain at a substantial number of about $20\%$ even in the outskirts of clusters at a distance of about $[0.7, 0.9] \mpch$.

In observations, we do not have access to the 3-D radius, but the projected radius. In Fig.~\ref{fig:orphan_fraction}, we compare the orphan fraction as a function of 3-D radius to that obtained as a function of the 2-D radius using green colour. The use of the projected radius mixes the lower orphan fraction outskirts and thus results in a comparatively lower fraction of orphan galaxies. The fraction drops to about 50 percent in the innermost $[0.1, 0.3] \mpch$ bin, and to about $20$ percent in the outermost $[0.7, 0.9]\mpch$ bin. These comparisons show the expectation in the ideal case.

However, observations of optically identified clusters do not select overdensities of galaxies in 3-D space and have to deal with interloper galaxies due to the imperfections of photometrically determined redshift for potential member galaxies. As we showed in the previous section our mock redMaPPer selection is able to mimic many of these effects present in the real redMaPPer cluster catalogue. With the member galaxies selected from such a catalogue, we have also computed the orphan fraction as a function of their projected radius from the assigned cluster center. These inferred values are shown with red colour points in Fig.~\ref{fig:orphan_fraction}. Surprisingly the orphan fraction trends mimic the orphan fraction as expected with the true 2-D projected orphan fractions, regardless of the presence of interloper galaxies.

Comparing these orphan fractions to the upper limits derived in our observational work, we see that the fraction of orphan galaxies is in 1-$\sigma$ agreement with our upper limits in the innermost regions of galaxy clusters. However if such trends remain true, then it implies that a substantial fraction of galaxies should have lost almost all of the dark matter mass associated with their subhaloes and do not leave much margin for the remaining galaxies to lose their subhalo masses due to incomplete stripping. The difference grows even larger as we consider satellite galaxies at farther distances away from the cluster center. The {\sc  UniverseMachine} shows a larger fraction of orphan galaxies than the upper limits from observations allow, even after we account for the errors on the upper limits themselves. We have further tested that our conclusions are robust to small changes to the threshold values of $M^*$ and SSFR values in our analysis, or to the use of different cosmological volumes to which the {\sc  UniverseMachine} model has been applied.

Although such large fractions of orphan population allow the {\sc  UniverseMachine} to simultaneously fit the clustering of faint galaxies and at smaller distances, the resultant orphan fractions in galaxy clusters would imply a large fraction of the galaxies have most of their dark matter haloes around them which are lost. One could argue whether such differences arise because of the artificial disruption of subhaloes in numerical simulations. The {\sc  UniverseMachine} identifies galaxies associated with subhaloes that fall below its detection limits as orphans. After this point they evolve the mass of the subhalo using an average mass loss prescription and with an orbit governed by a simplified model for the density profile of the halo. The mass loss process is assumed to be orbit independent.

It is worth noting that the loss of subhalo detectability is usually expected to be a larger effect in the interior regions of galaxy clusters, rather than in the outskirts. The difference in the outskirts would require a substantial fraction of subhaloes to have made at least one pericentric passage which would have made them vulnerable for artificial disruption due to numerical effects. Nevertheless, numerical effects are still likely to be the primary reason for the mismatch, since the {\sc  UniverseMachine} model is constrained using the cross-correlation of less massive galaxies with galaxies with $M_*>10^{11} \msunhh$ (considering them as proxy for galaxy clusters). If the orphan galaxies were truly overestimated then the cross-correlation of less massive galaxies with such more massive galaxies would have been larger than that seen in observations. Although we note that there is a large scatter in halo masses at high stellar masses, and it might be useful to compare cross-correlations of less massive galaxies in galaxy clusters as direct constraints. Improvements to the algorithm to find and track subhaloes by looking at the most bound particles in a subhalo before it fell in to the halo can help resolve numerical issues. Subhalo finders such as HBT+ \citep{Han2017} or Symfind \citep{Mansfield2023} use individual particle data in multiple snapshots and thus can identify and track subhaloes well past the point that they become unresolvable in single snapshot subhalo finders. It would be useful to see if the {\sc  UniverseMachine} algorithm run on halo(subhalo) catalogues found by such algorithms is able to satisfy the upper limits on orphan fractions.

We also observe that the masses of subhaloes assigned to orphan galaxies by the {\sc  UniverseMachine} shows no cluster-centric distance dependence of the mass difference of such subhaloes compared to equivalent galaxies in the field. This is likely a result of the simplified model used to treat subhaloes that get lost due to numerical issues. Given the treatment of such subhaloes, it is also not possible to track the actual mass around these subhaloes using their positions given in the {\sc  UniverseMachine}. The inclusion of observational constraints such as the weak lensing signals around subhaloes may help constrain the {\sc  UniverseMachine} model for satellite galaxies better and will be important to resolve this apparent mismatch.

Another direct empirical way to explore the existence of such a large fraction of orphan galaxies is to obtain the true tangential shear around satellite galaxies. This will involve a deconvolution of the shape measurement errors and the intrinsic ellipticities of background galaxies (shape noise) from the distribution of the observed tangential shear on background galaxies around satellite galaxies. Such a deconvolution should result in a bimodal distribution of tangential shears, one peak with a signal corresponding at most to the baryonic mass of the satellite galaxy, while the other corresponding to the shears around galaxies which still have substantial subhaloes present around them. We intend to explore these and other possibilities in future work.

\section{Summary}

We presented a comparison of the orphan galaxy prevalence in the empirical galaxy formation model {\sc  UniverseMachine} with the observational upper limits on orphan fraction obtained in \citetalias{Kumar2022} using the Subaru HSC weak lensing signal of satellite galaxies from the SDSS redMaPPer catalogue. The orphan fraction limits from observations can be affected by interloper galaxies which masquerade as satellites in observations. Therefore, we ran a mock redMaPPer algorithm on the {\sc  UniverseMachine} galaxy catalogues to obtain a catalogue of galaxy clusters and their members as would be found in real data. We showed that the properties of our mock redMaPPer catalogue mimic those of the original SDSS redMaPPer catalogue. In particular, our catalogue appropriately mimics the underlying mass-richness relation of galaxy clusters. We proceeded to investigate the dependence of the prevalence of orphan galaxies as a function of the cluster-centric distance in such optically selected galaxy clusters. Our results can be summarized as follows.
\begin{enumerate}
    \item The {\sc  UniverseMachine} galaxy catalogue shows a continuous decline in the prevalence of orphan galaxies as the separation of satellites increases from the cluster center ranging from 60 percent at projected separations between $[0.1, 0.3] \mpch$ to about 20 percent between $[0.7, 0.9]\mpch$.
    \item Projection effects in galaxy clusters lead to a difference in orphan prevalence when measured in 3-D vs 2-D proxy of separation of the satellites from the cluster center. However, this difference diminishes for the satellites present at larger separations from the cluster center. The fraction of orphan galaxies obtained from the mock redMaPPer catalogue of {\sc  UniverseMachine} galaxies closely mimic results of orphan fractions of true satellite member galaxies when using their projected separations from the cluster center. This shows that the orphan fraction constraints may not be significantly affected by the presence of interloper galaxies.
    \item The orphan fractions in the mock redMaPPer catalogue are consistent with the upper limits on orphan fractions from \citetalias{Kumar2022} for satellites with projected separations of $[0.1, 0.3] \mpch$. However for larger separations, we find that the fraction of orphan galaxies in the {\sc  UniverseMachine} are significantly higher than the upper limits established in \citetalias{Kumar2022}.
\end{enumerate}
We suggest that this discrepancy could be a result of the premature loss in the tracking of subhaloes in the halo catalogues which are based on the {\sc Rockstar} algorithm, as well as the subsequent simplified orbit-independent treatment in the {\sc  UniverseMachine} for the subsequent average mass loss in such subhaloes. Both these treatments can be improved by using the constraints from the weak lensing of satellite galaxies while constraining the {\sc  UniverseMachine} model.

\section*{Acknowledgements}
The authors would like to thank Peter Behroozi, Navin Chaurasiya, Divya Rana, Priyanka Gawade, Susmita Adhikari, Arka Banerjee, Souradip Bhattacharyya, Preetish K Mishra, Aseem Paranjape and Masahiro Takada for useful discussions and suggestions throughout this work and on the draft version of this manuscript.

\bibliographystyle{mnras}
\bibliography{bibliography}

\bsp	

\label{lastpage}
\end{document}